# Modeling Common Cause Failure in Dynamic PRA

Claudia Picoco[a], Valentin Rychkov[a]

[a]EDF R&D, Palaiseau, France

**Abstract**

In this paper we propose a dynamic model of Common Cause Failures (CCF) that allows to generate common cause events in time. The proposed model is a generalization of Binomial Failure Rate Model (Atwood model) that can generate staggered failures of multiple components due to a common cause. We implement the model using statechart formalism, a similar implementation can be adopted in other modeling languages like Petri Nets or Hybrid Stochastic Automata. The presented model was integrated in a Dynamic PRA study.

## 1. Introduction

Common Cause Failures (CCF) are an important part of the Probabilistic Risk Assessment (PRA) of nuclear power plant. They often represent important contributors to the overall risk appearing in the first ranks of minimal cut sets.

NRC defines CCF in the following way : "*A CCF event consists of component failures that meet four criteria: (1) two or more individual components fail, are degraded (including failures during demand or in-service testing), or have deficiencies that would result in component failures if a demand signal had been received, (2) components fail within a selected period of time such that success of the probabilistic risk assessment (PRA) mission would be uncertain, (3) components fail because of a single shared cause and coupling mechanism, and (4) components fail within the established component boundary.*" (NRC, 2007)

When a stochastic event (e.g., a failure) is modeled in a PRA, the following steps are performed (IAEA, 2010):
1) An assumption about the timing of the event is made based on the engineering knowledge about system behavior. Often the choice is made to consider that the failure occurs at the beginning of the scenario.
2) The scenario consequence is evaluated based on the timing of the event a support study is performed.
3) The PRA model is built based on the results of the support study. For the considered scenario, the failure on demand as well as the failure to run are both considered to occur at the beginning of the scenario, and, therefore, modeled within the same sequence. This approach is often backed up by the argument that the failure of the components at the beginning of the scenario represents a bounding case that englobes the consequences of all the scenarios where failures occur in later times.

In general practice, the CCF modeling follows the same logic, as the unavailability of multiple redundant components at the beginning of the scenario is considered as the bounding scenario. Nevertheless, CCF of real systems do not necessarily occur simultaneously. They may occur at the same time, for example because of a hazard and spatial proximity, but they may also appear in a staggered way. They are defined as failures of similar equipment occurring within a limited timespan. Therefore, some situations may require a refined approach to represent stochastic failures in time.

In these cases, dynamic methods for probabilistic safety assessment can be used for explicitly taking into account the time of occurrence of failure events into the risk quantification framework (Aldemir, 2013).

In dynamic PRA the time distribution of independent failures is, in general, modeled using exponential distribution. Nevertheless, the modeling of time distribution of failures due to CCF event is much less of a common practice.

Up to date, a limited number of studies, such as (Deleuze et al., 2016) and (Donat.et al., 2015) discussed the topic. (Donat.et al., 2015) proposed a modeling approach for CCF in a dynamic model that generate simultaneous failures of multiple components in time. (Deleuze et al., 2016) presented the possibility to model a staggered (and therefore more realistic) appearance of CCF but without a connection to a dynamic PRA.

In this paper, we present a modeling framework to represent CCF in dynamic models by using the Atwood model. This model was used in an industrial study (Rychkov et al., 2023) using statechart formalism. The proposed framework be adopted to different modeling approaches, if needed.

The paper is organized as follows. In Section 2 we recall the CCF theory model with a particular focus on Atwood model. In Section 3, we explain the modeling approach to model CCF in a dynamic approach via an example. For this we use the statechart formalism (Rychkov et al., 2023). In Section 4, we present the simulation results. Finally, in Section 5, we draw conclusions perspectives.

## 2. CCF Theory Model

Different CCF theory model exists such as the Alpha-factor (Mosleh et al., 1988), Multiple Greek Letters (Mosleh et al., 1988), Atwood (Atwood, 1986), that are currently used in standard PRA models. Nevertheless, there's no standard practice concerning CCF modeling within dynamic PRA study. One implementation we find in the literature concerns the use of Alpha factor model for the BDMP 4.1 knowledge base of KB3 (Donat et al., 2015). This approach indeed allows to integrate CCF events in a dynamic model, but it produces only simultaneous failures of multiple components due to a CCF event.

As mentioned in the introduction, one of the main challenges of modeling CCF in a dynamic model is that, as it happens in the reality, after the occurrence of the CCF event, failures may appear not simultaneously but in a staggered way.

In order to model this, we use a shock model such as the Atwood model. This model :
- have not the concept of CCF group, which minimizes its implementation into a dynamic model,
- easily allows the implementation of the occurrence of staggered failures following a CCF event (Deleuze et al., 2016).

### 2.1. A generalization of the Binomial Failure Rate model: the Atwood model

The Atwood model (Atwood, 1986) is a generalization of the Binomial Failure Rate (BFR) model (Atwood, 1986). As described in (Nguyen, 2019) concerning the Atwood model :

*"The main idea is to model component failures in terms of individual item failures and outside shocks that affect the survival status of all the components in a system. Individual components failures are described using independent and identically distributed exponential lifetimes with common failure rate, λ. Other parameters in the first version of BFR model are µ, the rate of CCF, and p, the probability that each component fails because of outside common cause shocks. When an outside shock occurs, it is assumed that each component has the same probability to fail, independently from each other [...] an independent lethal shock modeled by a Poisson distribution with a rate omega allows to cover additional situations."*

The Atwood model is a shock CCF model that supposes the possibility of the occurrence of (i) a lethal shock event (that will impact all the components) and (ii) a non-lethal shock event (that will impact partial number of components). The model makes use of 4 parameters:
- $\omega$ : occurrence rate of lethal shocks ;
- $\mu$ : occurrence rate of non-lethal shocks ;
- $\rho$ : conditional probability of a component failure given the occurrence of non-lethal shock ;
- $\lambda_{ind}$ : independent failure rate.

In summary, during a given sequence :
- The component can fail independently with a failure rate $\lambda_{ind}$,
- A lethal shock can occur with an occurrence rate $\omega$, after which all components fail in a staggered way.
- A non-lethal shock can occur with an occurrence rate $\rho$, after which some components fail in a staggered way.

From an implementation point of view, in order to guarantee the staggered appearance of the failures, this translates in :
- The independent failure is sampled from an exponential distribution of parameter $\lambda_{ind}$,
- The occurrence of the lethal shock is sampled from an exponential distribution of parameter $\omega$. If this happens, the failure of the component is resampled from a truncated exponential distribution with a new parameter that is defined such that the probability to fail within the mission time is equal 1,

- The occurrence of the non-lethal shock is sampled from an exponential distribution of parameter $\mu$. If this happens, the failure time of the component is resampled from an exponential distribution with a new parameter that is defined such that the probability to fail within the mission time is equal $\rho$.

## 3. The implementation in a dynamic study : an example

In order to illustrate the implementation of the modeling framework we use the example of 4 Emergency Diesel Generators (EDGs) working on 4 different trains using statechart[1] formalism (Harel, 1987). Statecharts are used by numerous methods that target diverse application domains such as robotics, internet agents, simulation, health and safety (Spanoudakis, 2021).

CCF model is implemented in a dynamic model (Rychkov et al., 203) with itemis CREATE tool (itemis, 2024).

### 3.1. CCF Simulation Model

The model consists of two parts :
- CCF event generator, in charge of sampling and tracking the occurrence of the lethal or non-lethal shock.
- EDGs model, in charge of modeling the behavior of the EDGs. A model of the EDG is implemented and the replicated 4 times, one representing each EDG.

*3.1.1. CCF event generator*

As mentioned earlier, the task of the CCF event generator is to sample and track the occurrence of lethal and non-lethal shock events.

Fig. 1 represents the statechart model of the CCF generator. The sampling of the time of occurrence of the lethal and non-lethal shock, *tf_Lethal* and *tf_NonLethal* respectively, is performed through a *tf* C++ function with parameter *omega* and *mu* respectively. The *tf* function implement the sampling from an exponential distribution. After the sampling, the **Idle** state is reached. If the non-lethal shock occurs first, then the transition "after *tf_NonLethal* s" is taken, evolving the statemachine toward the state **NonLethal_failed**. This will correspond to an event trigger "CCF_NonLethal". If, instead, the lethal shock occurs first, then the transition "after *tf_Lethal* s" is taken the state **Lethal_failed** is entered. This will correspond to the event "CCF_Lethal" triggered. In the presented model we assume that it is not possible to have multiple shocks within the same sequence.

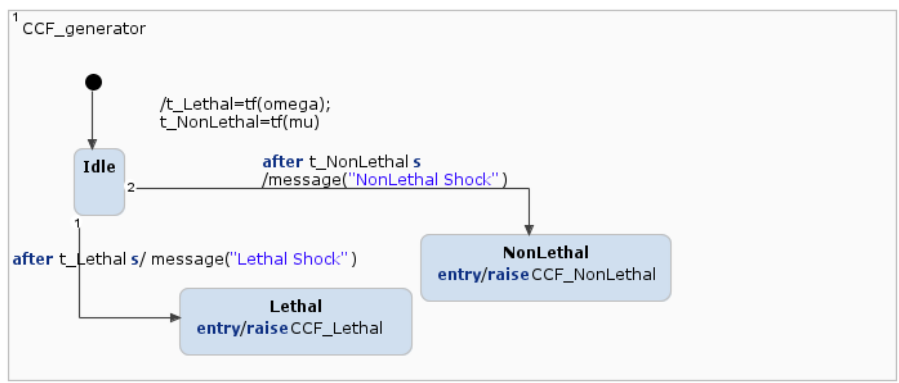

Fig. 1. Statechart model of the CCF event generator

---

[1] A state machine is a behavior model. It consists of a finite number of states and is therefore also called finite-state machine (FSM). Based on the current state and a given input the machine performs state transitions and produces outputs. (itemis, 2024)

*3.1.2. EDG model*

The EDG model is replicated 4 times to represent each EDG. Fig. 2 shows the example of EDG1.

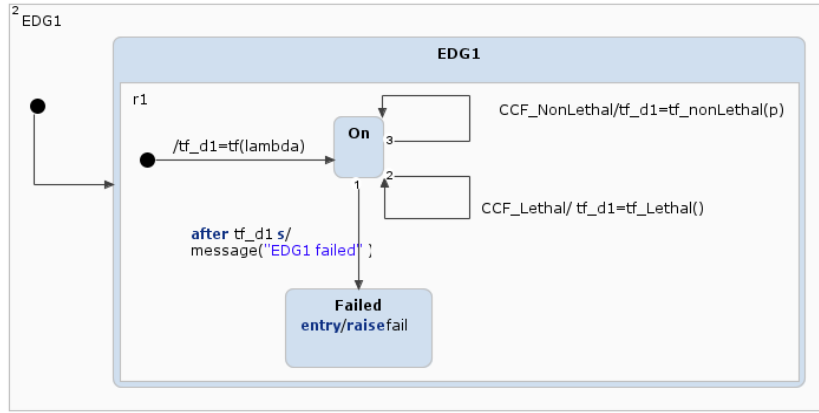

Fig. 2. Statechart model of the EDG component (example of EDG1)

At the beginning of the simulation, failure time *tf_d1* of EDG1 is sampled from an exponential distribution with parameter lambda (= $\lambda_{ind}$) using the C++ function *tf* :
- If the independent failure occurs first, then the transition "after *tf1* s" is taken, evolving the statemachine to the state **Failed** state.
- If the lethal shock occurs first (represented by the CCF_Lethal event triggered by the CCF event generator) then *tf_d1* is resampled by the function tf_Lethal. This function resampled *tf_d1* from a truncated exponential distribution with a parameter *lambda_eff_Lethal* that depends on the remaining time until mission time. This *lambda_eff_Lethal* is defined such that the failure conditional probability within timespan is 1.
- If the non-lethal shock occurs first (represented by the CCF_NonLethal event triggered by the CCF event generator) then *tf_d1* is resampled by the function *tf_NonLethal*. This function resampled *tf_d1* from an exponential distribution with a parameter *lambda_eff_NonLethal* that depends on the remaining time until mission time. This is defined such that the failure conditional probability within timespan is $\rho$.

*3.1.3. The overall model*

The overall model consists of 4 EDGs and 1 CCF generator as shown in Fig. 3.

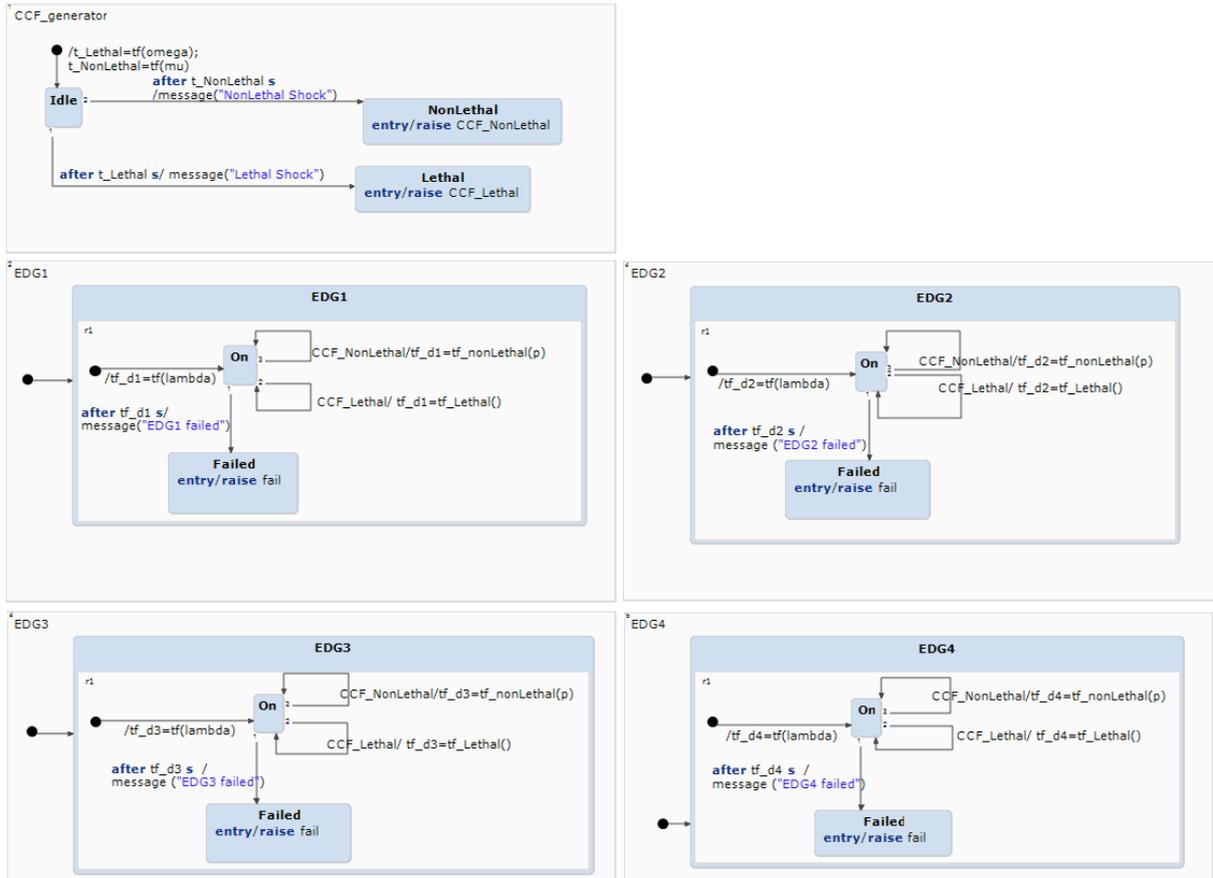

Fig. 3. The overall Model

The original reliability data from the PRA model are shown in Table 1.

Table 1. Reliability Data (INL, 2020)

| | |
|---|---|
| $\alpha_1$ | 9,87E-01 |
| $\alpha_2$ | 7,06E-03 |
| $\alpha_3$ | 4,55E-03 |
| $\alpha_4$ | 1,54E-03 |
| $\lambda$ | 1,18E-03 |

## 4. Verification

In order to verify the model, we decide to start with the alpha model parameters used in the PRA model. Then, convert the alpha model parameters in Atwood model parameters, for that we solved the 4 equations with 4 unknowns' linear system. The corresponding Atwood parameters, resulting from the conversion, are shown in Table 2.

Table 2. Alpha model to Atwood model conversion

| | | | | |
|---|---|---|---|---|
| $\alpha_2$ | 7,06E-03 | | $\omega$ | 2,04E-06 |
| $\alpha_3$ | 4,55E-03 | | $\mu$ | 8,71E-05 |
| $\alpha_4$ | 1,54E-03 | $\Leftrightarrow$ | $\rho$ | 4,92E-01 |
| $\lambda_{tot}$ | 1,18E-03 | | $\lambda_{ind}$ | 1,14E-03 |
| $\alpha_1$ | 9,87E-01 | | | |

| | |
|---|---|
| $\alpha_{tot}$ | 1,02 |

Next, the model is simulated $1^{E}7$ times via a MonteCarlo simulations over 24 hours. Fig.4-7 illustrate examples of possible outcomes of simulated sequences represented as sequence diagrams.

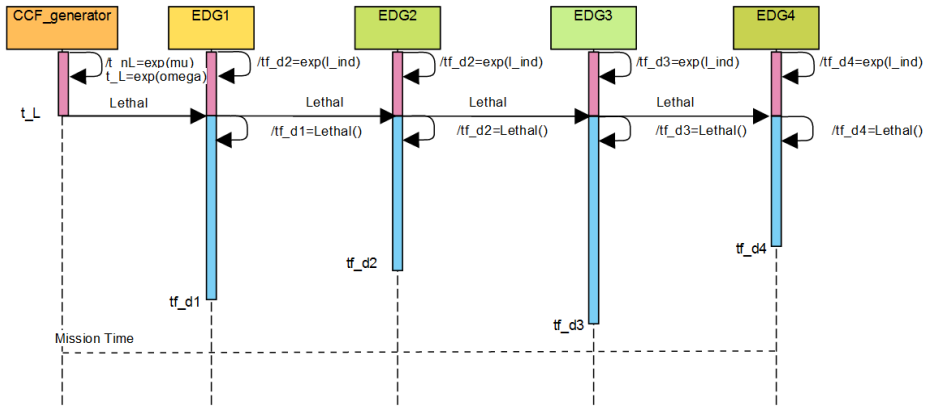

Fig. 4 A common cause failure of the order of 4 due to a lethal shock. The CCF generator generates a lethal shock at time t_L, that provokes resampling of the failure times of each EDG using exponential distribution truncated at the mission time Then, each EDG fails independently at different time tf_d1,2,3,4

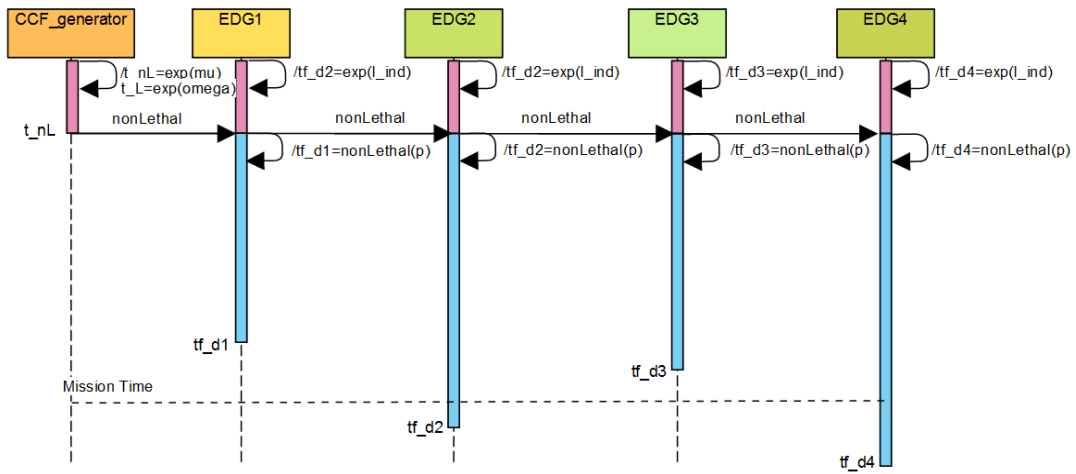

Fig. 5 A common cause failure of 2 components. The CCF generator generates a non-lethal shock at time t_nL, that provokes resampling of the failure times of each EDG using exponential distribution with the effective lambda calculated in such a way that the probability of failure within the mission time is equal to ρ. Then, each EDG fails independently at different time tf_d1,2,3,4. Failures of EDG 1 and 3 fall inside the mission time, thus only these failures are accounted for.

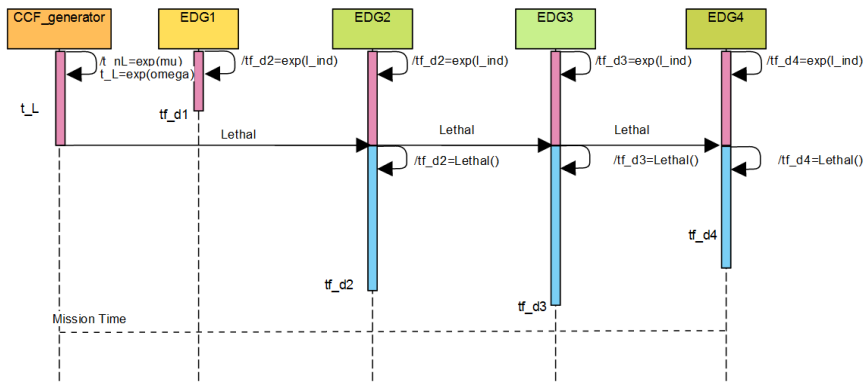

Fig. 6 An independent failure of 1 component and common cause failure of 3 components. An EDG1 fails due to independent cause at tf_d1, then the CCF generator generates a lethal shock at time t_L, that provokes resampling of the failure times of EDGs 2,3,4 using exponential distribution truncated at mission time. Then each EDG fails independently at different time tf_d2,3,4

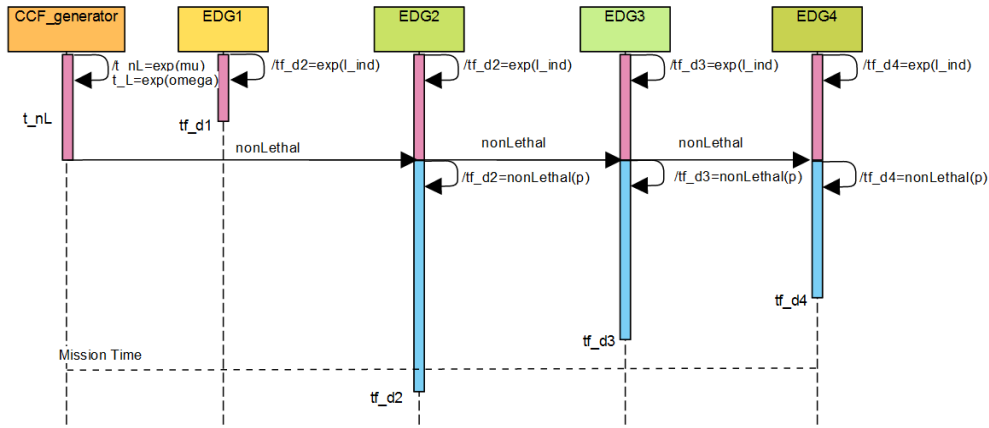

Fig. 7 An independent failure of 1 component and common cause failure of 2 components. EDG1 fails due to independent cause at tf_d1, then the CCF component generates a non-lethal shock at time t_nL, that provokes resampling of the failure times of EDGs 2,3,4 using exponential distribution with the effective lambda calculated in such a way that the probability of failure within the mission time is equal to ρ. Then each EDG fails independently at different time tf_d2,3,4, but failure of only EDG3 and 4 falls inside the mission time thus only accounted for.

Results are then analyzed to:
- Estimate the Atwood parameters (in order to verify our program model)
- Estimate the alpha parameters (in order to verify our Atwood → Alpha conversion).

The estimation of the alpha model and Atwood model parameters are here shown:

$$\omega = \frac{\text{number of simulated lethal shocks}}{\text{mission time} * \text{number of simulations}} = 2.14\text{E} - 06/h \tag{1}$$

$$\mu = \frac{\text{number of simulated non lethal shocks}}{\text{mission time} * \text{number of simulations}} = 8.66\text{E} - 05/h \tag{2}$$

$$\rho = \text{mean number of failures per non lethal shock per component} =$$
$$\frac{\text{number of failures in non lethal shocks}}{4 * \text{number of non-lethal shocks}} = 0.503 \tag{3}$$

These BFR parameters imply the following estimation of Alpha model parameters.

$$\alpha_2 = \frac{\text{number of simulation with two failed components observed}}{\text{totale number of observed failures}} = 6.96\text{E} - 3 \tag{4}$$

$$\alpha_3 = \frac{\text{number of simulation with three failed components observed}}{\text{totale number of observed failures}} = 4.41\text{E} - 3 \tag{5}$$

$$\alpha_4 = \frac{\text{number of simulation with four failed components observed}}{\text{totale number of observed failures}} = 1.51\text{E} - 3 \tag{6}$$

Eventually, as a further confirmation, we also estimated the parameter $\lambda$:

$$\lambda_{\text{tot}} = \frac{\text{total number of observed failure}}{4 * \text{mission time} * \text{number of simulations}} = 1.17\text{E} - 3/h \tag{7}$$

Table 3 presents an overview of calculations and results. We find a good agreement between original and estimated alpha and Atwood parameters.

Table 3. Overview of parameters estimation

|  | Estimated | Input |
|---|---|---|
| $\alpha_2$ | 6,97E-03 | 7,06E-03 |
| $\alpha_3$ | 4,40E-03 | 4,55E-03 |
| $\alpha_4$ | 1,52E-03 | 1,54E-03 |
| $\lambda_{tot}$ | 1,18E-03 | 1,14E-03 |
|  |  |  |
| $\omega$ | 2,14E-06 | 2,04E-06 |
| $\mu$ | 8,60E-05 | 8,71E-05 |
| $\rho$ | 0,51 | 0,49 |

## 5. Conclusions and Perspectives

This paper presents a first approach to model dynamically CCF using a shock model. The model has been applied to a real case (Rychkov et al., 2023). Nevertheless, some open questions concerning how to model CCF in time remain and further research should be performed in order to :
- Validate the Atwood model for the case of CCF group of different sizes;
- Validate the proposed staggered mechanism for longer mission times (which may be affected by the lack of operating data);
- Investigate how to integrate CCF on demand;
- Explore how to model CCF in case of repair;
- Compare parametric model (Alpha) vs shock models (Atwood) for dynamic studies.